\documentclass[useAMS]{mn2e}
\usepackage{times}

\def\eq#1 {eq.~(\ref{eq:#1})}

\def\etal{{\it et al.\ }}

\def\eg{{\it e.g.},}
\def\ie{{\it i.e.},}

\def\ltsima{$\; \buildrel < \over \sim \;$}
\def\lsim{\lower.5ex\hbox{\ltsima}}
\def\gtsima{$\; \buildrel > \over \sim \;$}
\def\gsim{\lower.5ex\hbox{\gtsima}}
\def\ga{\mathrel{\hbox{\rlap{\hbox{\lower4pt\hbox{$\sim$}}}\hbox{$>$}}}}
\def\la{\mathrel{\hbox{\rlap{\hbox{\lower4pt\hbox{$\sim$}}}\hbox{$<$}}}}

\def\la{\dangler}

\def\pmb#1{\setbox0=\hbox{#1}%
 \kern-.025em\copy0\kern-\wd0
 \kern.05em\copy0\kern-\wd0
 \kern-.025em\raise.0433em\box0}

\def\3hmpc{\, ( h^{-1} {\rm Mpc})^3}

\def\r0p { r{_0^\prime}}

\title[LOFAR as a Probe of Cosmological Reionisation Sources]{LOFAR as
a Probe of the Sources of Cosmological Reionisation} \author[Zaroubi
\& Silk]{Saleem Zaroubi$^1$ \& Joseph Silk$^2$ \\$^1$Kapteyn
Astronomical Institute, University of Groningen, Landleven 12, 9747 AG
Groningen, The Netherlands\\$^2$Astrophysics Department, University of
Oxford, Keble Road, Oxford OX1 3RH }

\begin{document}

\maketitle
\begin{abstract}
We propose use of the thickness of the ionisation front as a
discriminant between alternative modes of reionisation in the early
universe, by stars or by miniquasars. Assuming a
photoionisation-recombination balance, we find that for miniquasar
sources the transition from neutral to ionised intergalactic medium is
extended and has two features.  The first is a sudden steep increase
in the neutral fraction with a typical width of $5-10$ comoving
megaparsecs, depending on the miniquasar power. The second feature is
a long wing that represents a much slower transition from neutral
fraction of $\approx 0.8$ to $1$.  The angular resolution of LOFAR is
expected to resolve these scales and will, therefore, play an
important role in discriminating the hard sources of ionising photons
from the stellar ones.
\end{abstract}


\section{Introduction}
\label{introduction}
The universe is generally considered to have become reionised at a
redshift larger than 10. This result, coming from analysis of the WMAP
polarisation power spectrum (Kogut \etal 2003 \& Spergel
\etal\ 2003), came as a surprise, in the context of earlier studies of
the Gunn-Peterson effect inferred to be present at $z\sim 6$ (\eg\
Fan, X., \etal\ 2002) and of the high temperature of the intergalactic
medium at $z\sim 3$ (Theuns \etal\ 2002; Hui \& Haiman 2003). Given
this result, the ionising sources cannot be known quasars or normal
galaxies.  Rather, recourse must be had to Population III stars or to
miniquasars, both of which represent hypothetical but plausible
populations of the first objects in the universe, and that are
significant sources of ionising photons.

The impact of stellar sources have been studied by many authors (\eg\
Cen 2003, Ciardi, Ferrara, \& White 2003, Haiman \& Holder 2003,
Wyithe \& Loeb 2003, Sokasian \etal 2003, Somerville \& Livio
2003). These studies find in general that in order to provide enough
ionising flux at or before $z=15,$ for the usual scale-invariant
primordial density fluctuation power spectrum, one needs Population
III stars, which provide about 20 times more ionising photons per
baryon than Population II (Schaerer 2002, Bromm, Kudritzki, \& Loeb
2001), or an IMF that initially is dominated by high mass stars
(Daigne \etal\ 2004). This is in agreement with recent numerical
simulations of the formation of first stars from primordial molecular
clouds that suggest that the first metal-free stars were predominantly
very massive, $m_*\gsim 100 M_\odot$ (Abel, Bryan, \& Norman 2000;
2002 \& Bromm, Coppi \& Larson 2002).

Theoretically, there is some tension between the low amplitude of the
fluctuations as interpreted from low redshift data (\eg\ Croft \etal
2002, Zaroubi \etal 2004) and the early ionisation observed by the
WMAP satellite. Most of the theoretical and numerical results
mentioned earlier require a somewhat higher value of the fluctuation
amplitude and would have a harder time satisfying all of the
observational constraints with lower amplitude ($\sigma_8\approx
0.8$). It is unclear whether this tension is caused by the theoretical
uncertainties in the physical details of the ionisation process or it
is an indication of the existence of a non-standard cosmological model
in which nonlinear structures assemble earlier than in the normal
$\Lambda$CDM scenario (\eg\ Sugiyama, Zaroubi \& Silk 2004; Avelino \&
Liddle 2004).

Miniquasars have also been considered as a significant ionising source
(\eg\ Ricotti \& Ostriker 2004a; 2004b, Madau \etal 2004, Oh 2000,
2001, Dijkstra, Haiman \& Loeb 2004). These latter are as plausible as
Population III stars, whose nucleosynthetic traces have not yet been
seen even in the most metal-poor halo stars nor in the high $z$ Lyman
alpha forest, in view of the correlation between central black hole
mass and spheroid velocity dispersion (Ferrarese 2002, Gebhardt \etal
2000). This correlation demonstrates that seed black holes must have
been present before spheroid formation. Indeed, recent observations of
a quasar host galaxy at $z=6.42 $ (Walter et al. 2004) (and other
quasars) suggest that supermassive black holes were in place and
predated the formation of the spheroid. Theory suggests that the seeds
from which the SMBH formed amounted to at least $1000 \rm M_\odot$ and
were in place before $z\sim 10$ (Islam, Taylor \& Silk 2003, Madau \&
Rees 2001).

Can we distinguish between the alternative hypotheses of stellar
versus miniquasar ionisation sources? One distinguishing feature is
the intrinsic source spectrum, which is thermal for stars but with a
cut-off at a few times the Lyman limit frequency, but power-law for
miniquasars with a spectrum that extends to high energies with nearly
equal logarithmic increments in energy per decade of frequency.  We
show here that there is a dramatic difference between these two cases
in the widths of the ionisation fronts. Only the miniquasar model
translates to scale-dependent 21 cm brightness temperature
fluctuations that should be measurable by LOFAR.

\section{Width of x-ray ionisation fronts}
\label{Theory}

Typically, the spectrum of the UV radiation of stellar sources is
roughly flat with a cutoff at a few times the Lyman limit
frequency. This results in a very sharp transition from ionised to
neutral IGM. The thickness of such a region may be estimated from the
mean free path of a photon with energy $E$ within a neutral medium,
\begin{equation}
\langle l_E \rangle \approx \frac{1}{n_H \sigma_H(E)},
\end{equation} 
where $n_H\approx 2.2\times10^{-7}~\mathrm{cm^{-3}} (1+z)^3$ (Bennett
\etal 2003) is the mean number density of hydrogen at a given
redshift, and $\sigma_H(E) = \sigma_0 \left({E_0/E}\right)^3$ is the
bound-free absorption cross-section for hydrogen with
$\sigma_0=6\times 10^{-18}~\mathrm{cm^2}$ and $E_0=13.6~eV$. At
$z=10,$ the mean free path for the most energetic photon from a
stellar source is of the order of $0.1$ comoving kpc.

In contrast, the power-law behavior of the radiation specific
intensity of miniquasars produces a much thicker ionisation front. In
the following calculation, we estimate this thickness assuming a
uniform mass density distribution in which a miniquasar is
embedded. This obviously does not take into account any mass or quasar
clustering properties; however, for the purpose of exploring the
thickness to within an order of magnitude and its qualitative
evolution with redshift and dependence on the black hole mass, this
assumption is acceptable. A more realistic numerical calculation will
be carried out in the future.

We are also going to assume ionisation-recombination equilibrium.
This may be safely assumed for the photons we are interested in as
they have a mean free path of $\approx 1-10$ comoving Mpc at redshift
$\approx 10-5$, which translates to a mean scattering time of about
0.5-5 million years, much less than a Hubble time at the relevant
redshift range.

The miniquasar source spectrum is taken to be
\begin{equation}
I(E)= A g E^{-0.8} e^{-E/E_c}~\mathrm{cm^{-2} s^{-1}}
\label{spectrum}
\end{equation}
where $g$ is the Gaunt factor and $A$ is normalized such that the
luminosity of the miniquasar is $10^{43}\mu~\mathrm{erg~s^{-1}},$ for
black holes of mass $10^5\mu \rm M_\odot,$ with $1\lsim \mu\lsim
10^3,$ and the cutoff energy $E_c$ is $200~\mathrm{keV}$ (Sazonov
\etal 2004).  This translates to a number of ionisations at a distance
$r$ from the source,
\begin{equation} 
{\cal N}(E;r) = e^{-\tau(E;r)} \frac{A
g}{\left(r^2/~\mathrm{Mpc^2}\right)} E^{-0.8}
\left(\frac{E}{E_i}\right) e^{-E/E_c},
\label{flux}
\end{equation}
with
\begin{equation} 
\tau(E;r)=\int_0^r n_H x_{HI} \sigma(E) dr.
\label{tau}
\end{equation}
Here $x_{H I}$ is the hydrogen neutral fraction.

The factor $E/E_i$ in equation~\ref{flux} is added to account for the
mean number of ionisations per ionising photon with $E_i \approx
36~\mathrm{eV}$.  This last choice is a simplifying approximation to
the number of ionisations per photon which has a more complex
dependence on energy and ionised fraction of hydrogen (Shull \& van
Steenberg 1985; Dijkstra, Haiman \& Loeb, 2004). The approximation we
adopt here is reasonable for photons within the energy range that
gives rise to the thickness of the ionisation front, \ie\ photons with
$E\approx 0.2-1~\mathrm{keV}$.
\begin{figure*}
\setlength{\unitlength}{1cm} \centering
\begin{picture}(16,10)
\put(-3.1, -3.5){\includegraphics{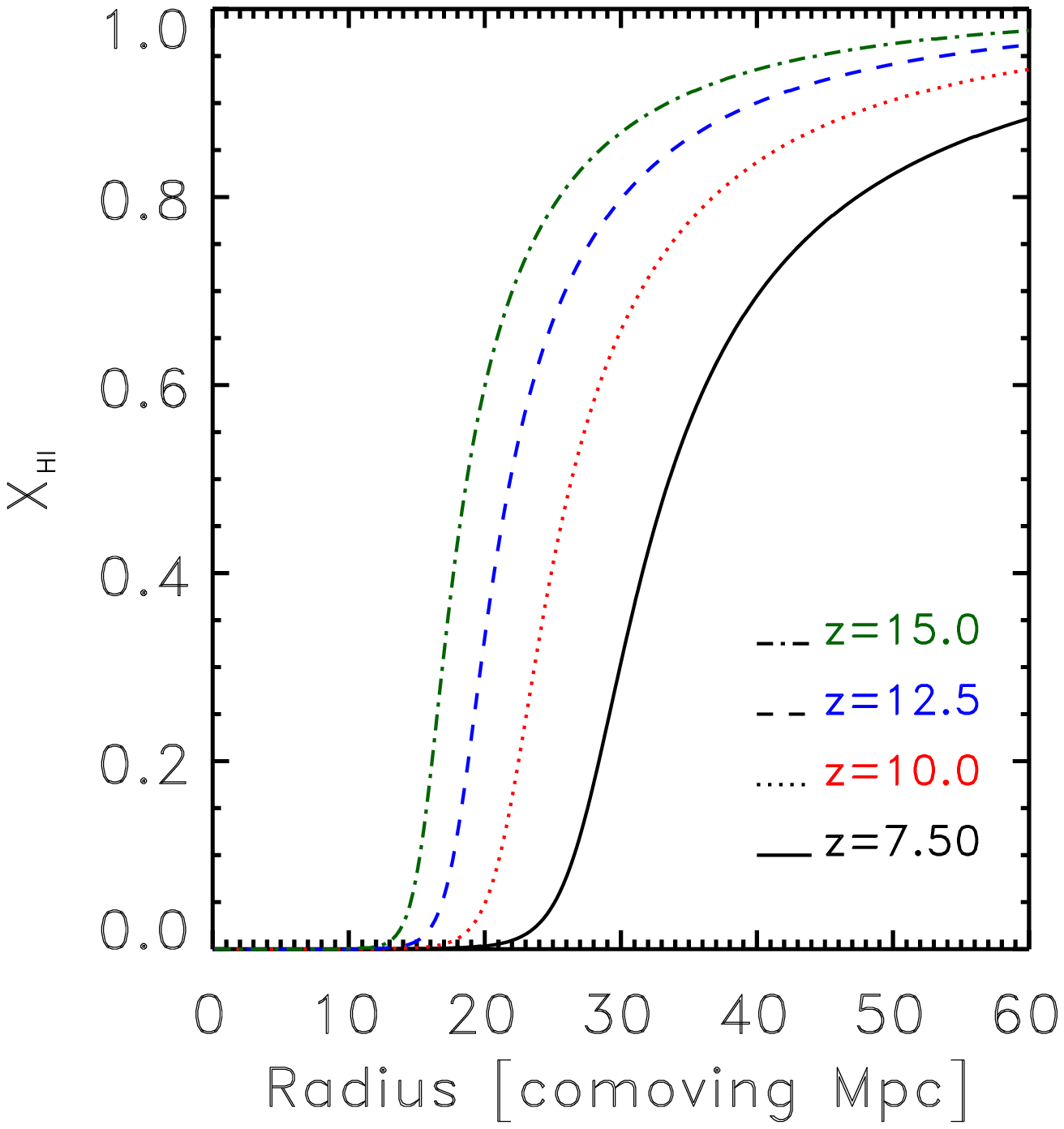}} \put(6.1, -3.5){\includegraphics{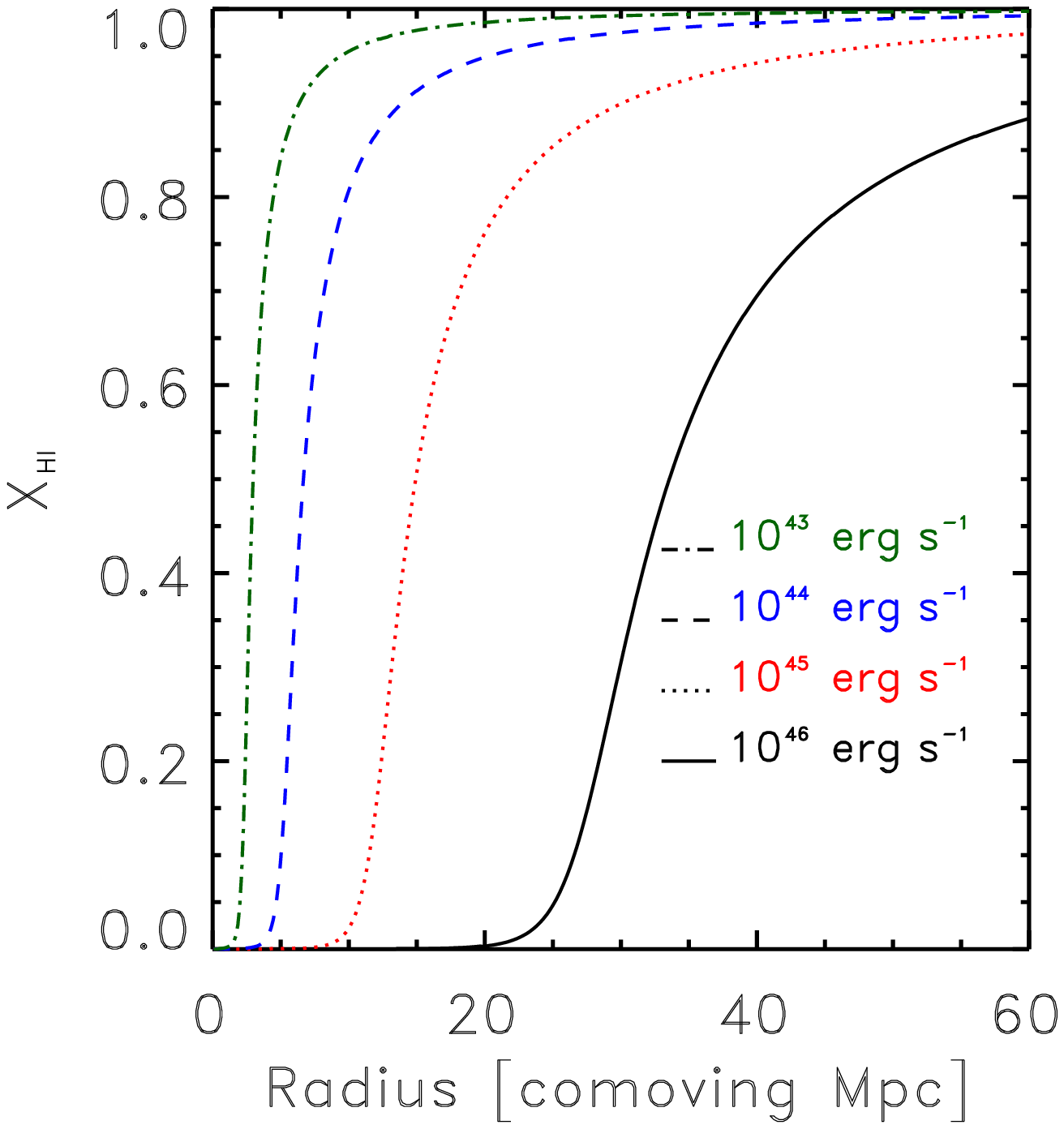}}
\end{picture}
\caption{The left panel shows the H I fraction as a function of
comoving distance from the source for redshift values of
$z=[7.5,10,12.5,15]$ for a source with $\mu = 10^3$ in units of the
Eddington luminosity for a $10^5~M_\odot$ black hole. The right panel is
similar to the left one except that here the redshift is fixed to
z=7.5 and the various curves reflect values of $\mu=1,10,100,1000$.  }
\label{fig:fig1}
\end{figure*}

The cross-section quoted earlier does not take into account the
presence of helium. In order to include the effect of helium, we
follow Silk \etal (1972) who modified the cross section to become
\begin{equation}
\sigma(E) = \sigma_H(E) +{n_{He} \over n_H} \sigma_{He} = \sigma_1
\left({E_0 \over E}\right)^3.
\end{equation}
A proper treatment of the effect of helium is accounted for by
defining $\sigma_1$ to be a step function at the two helium ionisation
energies corresponding to He I and He II. This however includes
lengthy calculations and complicates the treatment, and we therefore
choose $\sigma_1$ to be a smooth function of $E,$ an approximation
that will overestimate $\sigma(E)$ for low energy photons. For photons
with the relevant energy range, namely those responsible for creating
the thick transition layer, this is a reasonable assumption.

The equation of ionisation equilibrium which expresses the balance
between the rate of number density of ionisation and recombination
events is,
\begin{equation}
\alpha_{H I}^{(2)} n_H^2 (1-x_{H I})^2= \Gamma(E;r)~n_H x_{H I}
\left(1+{\sigma_{He}\over\sigma_{H} }{n_{He}\over n_{H}}\right)
\label{balance}
\end{equation}
Here $\Gamma(E;r)$ is the ionisation rate per hydrogen atom for a
given photon energy at distance $r$ from the source. Since we are
interested in the detailed structure of the ionisation front, $\Gamma$
is calculated separately for each $r$ using the expression,
\begin{eqnarray}
\Gamma (E;r) & = & \int_{E_0}^{\infty}\sigma(E) {\cal N}(E;r)
\frac{dE}{E} \\ & = & \frac{A g \sigma_1}{3
{\left(r^2/~\mathrm{Mpc^2}\right)}} \frac{E_0^{0.2}}{E_i}\times \\ & &
\times\int_0^1 y^{1/15} e^{-\tau_0(r) y -\frac{E_0}{E_c} y^{-1/3}} dy,
\end{eqnarray}
where we have introduced the following simplifying notation:
equation~\ref{tau} is written as $\tau(E;r) \equiv
\tau_0(r)~(E_0/E)^3$ where $\tau_0(r)$ is the optical depth for
photons with the Lyman limit frequency at a given $r$. We also changed
the integration variable to $y\equiv \left({E_0/E}\right)^3.$
$\alpha_{H I}^{(2)}$ is the recombination cross-section to the second
excited atomic level and has the values of $2.6 \times
10^{-13}T_4^{-0.85} \mathrm{cm^3 s^{-1}}$, with $T_4,$ the gas
temperature in units of $10000 K,$ assumed here to be unity.

Figure~\ref{fig:fig1} shows the solution of equation~\ref{balance} for
miniquasars with various luminosities and different redshifts. The
transition from neutral to ionised IGM is extended and has two
features. The first is a sudden steep increase in the neutral fraction
with a typical width of $5-10$ comoving megaparsecs (see left
panel). The second feature is a wing that represents a much slower
transition from $x_{H I}\approx 0.8$ to $1$. In the next section, we
will show that with LOFAR, one should be able to resolve both features
for the luminous miniquasars. For the less luminous quasars, one might
have a chance of just  detecting the wing feature with LOFAR.  Only a
radio telescope like SKA (the Square Kilometer Array) that has 10
times the total collecting area of LOFAR will be able to see the whole
transition down to miniquasars with $10^{43}\mathrm{erg~s^{-1}}$
luminosity.

\section{LOFAR predictions}

In radio astronomy, where the Rayleigh-Jeans law is usually
applicable, the radiation intensity, $I(\nu)$ is expressed in terms of
the brightness temperature, so that
\begin{equation}
I(\nu) = \frac{2 \nu^2}{c^2} k T_b,
\end{equation}
where $\nu$ is the radiation frequency, $c$ is the speed of light and
$k$ is Boltzmann's constant (Rybicki \& Lightman 1979). This in turn
can only be detected differentially as a deviation from $T_{CMB}$, the
cosmic microwave background temperature. The predicted differential
brightness temperature deviation from the cosmic microwave background
radiation is given by (Field 1958; 1959; Ciardi \& Madau 2003),
\begin{equation}
\delta T_b = 16~\mathrm{mK}~x_{HI} \left(1-\frac{T_{CMB}}{T_s}\right)
\left(\frac{\Omega_b h^2}{0.02}\right) \left[\left(\frac{1 +
z}{10}\right)\left(\frac{0.3}{\Omega_m}\right)\right]^{1/2}.
\end{equation}
Here $T_s$ is the spin temperature and $\Omega_m$ and $\Omega_b$ are
the mass and baryon density in units of critical density. In the
following calculation, we take $T_s$ to be significantly larger than
$T_{CMB}$. The enhancement in $T_s$ relative to $T_{CMB}$ could be
caused by the X-ray photons of the miniquasar itself or by an X-ray
background produced collectively by miniquasars formed at higher
redshifts (Ricotti \& Ostriker, 2004a; Nusser 2004, Tozzi \etal\
2000).  Furthermore, we are adopting a standard universe with a flat
geometry, $\Omega_bh =0.02$, $\Omega_m=0.3$ and $\Omega_\Lambda=0.7$.

Currently, there are a few experiments (\eg\ LOFAR, PAST and MWA) that
are being designed to directly measure $\delta T_b$ of the HI 21 cm
hyperfine line and probe the physics of the reionisation process by
observing the neutral fraction of the IGM as a function of
redshift. In this study, we focus on predictions for LOFAR, but our
conclusions could be easily applied to the other telescopes.

The LOFAR array consists geographically of a compact core area and 45
remote stations. Each remote station will be equipped with 100 High
Band antennas, 100 Low Band antennas. In the core area, with 2~km
diameter, there will probably be 32 substations. For the astronomy
application, there will be a total of 3200 High Band and 3200 Low Band
antennas in the core area. Currently, the planned maximum baseline
between the remote stations is roughly 100~km. The Low Band antenna
will be optimised for the 30-80 MHz range while the High Band antenna
will be optimised for the 120-240 MHz range.

The High Band antennas are sensitive enough to allow for the detection
of the brightness temperature produced by the high redshift $21$~cm
transitions.  In this band, LOFAR will have the ability to measure
brightness fluctuations as low as $5$ mK with spectral resolution of
$\approx 1~\mathrm{M Hz}$
and spatial resolution of about $3$~arcmin.  As currently configured,
LOFAR will be sensitive to the $21$~cm emission at redshift 6-11.5
over a field of 100 sq degrees.  First operation is foreseen towards
the end of 2006.

The resolution of LOFAR depends on the telescope baseline and observed
wavelength. As currently designed, the epoch of ionisation signal will
be sensitively measured mainly in the LOFAR core which has a FWHM
resolution of $\approx 3$ arcmin. For more information, see the LOFAR
web site: www.lofar.org

\begin{figure*}
\setlength{\unitlength}{1cm} \centering
\begin{picture}(16,10.)
\put(-3.1, -3.5){\includegraphics{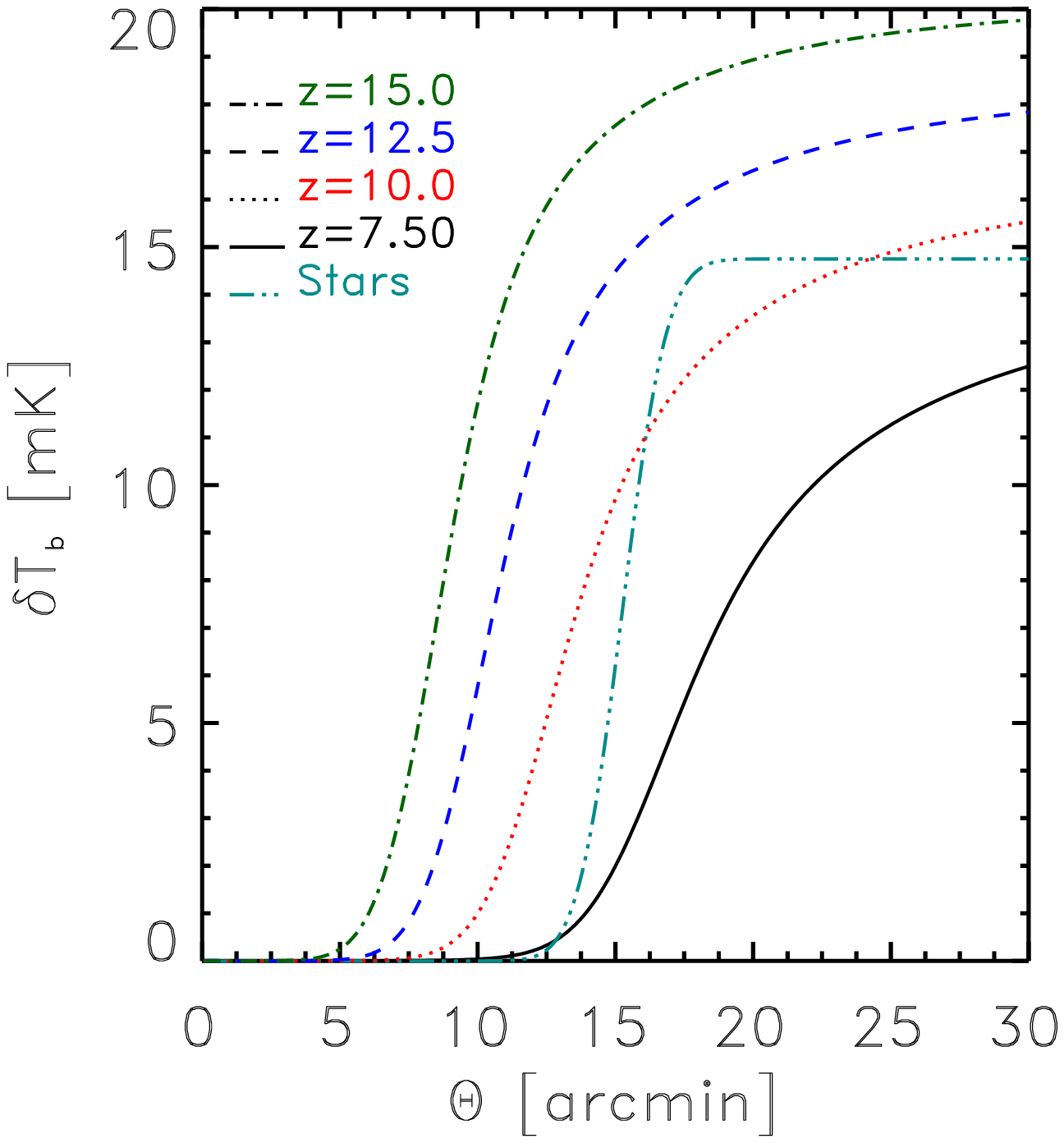}} \put(6.1, -3.5){\includegraphics{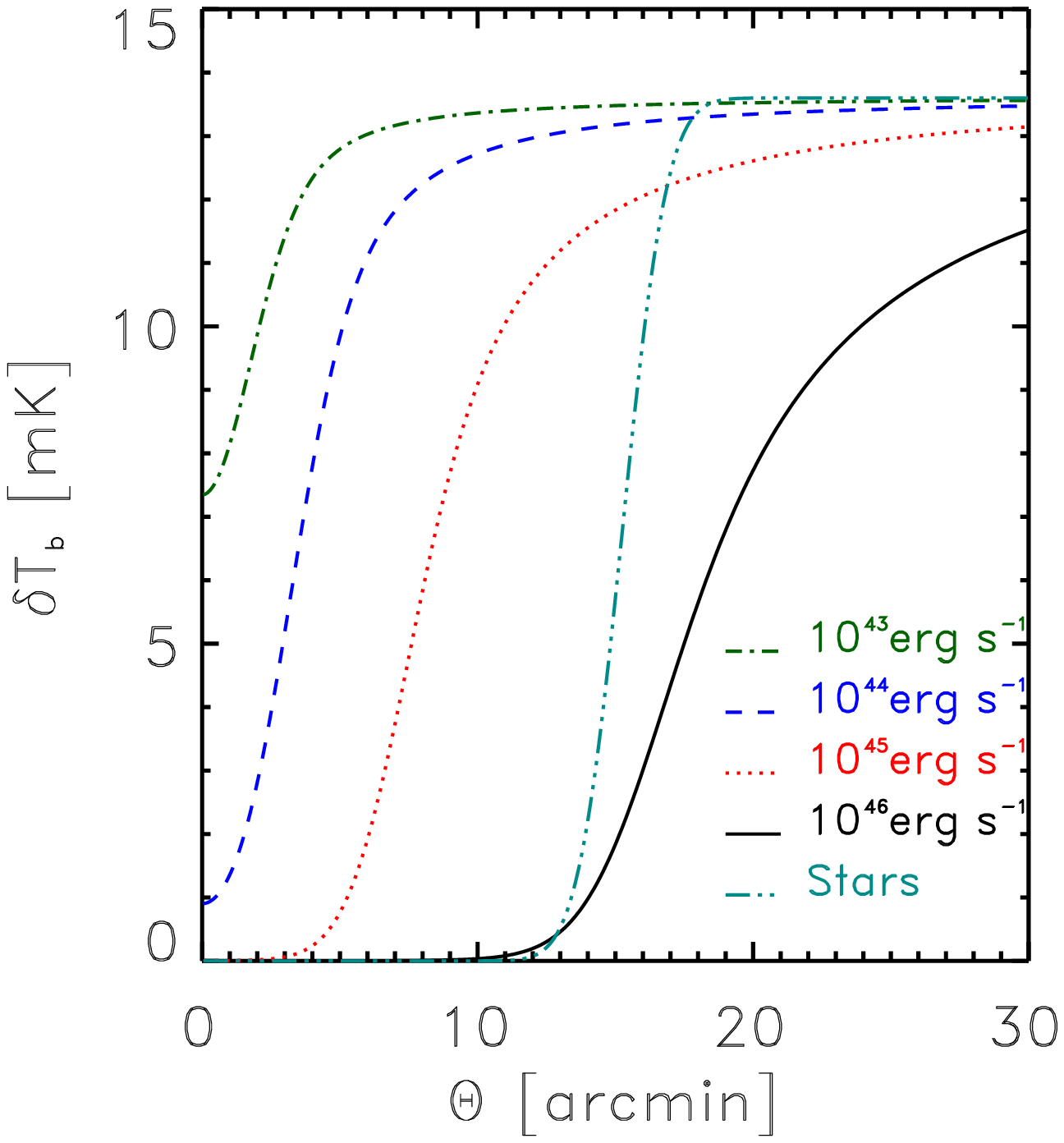}}
\end{picture}
\caption{The left panel shows $\delta T_b$ convolved with the 3 arcmin
FWHM LOFAR Gaussian beam as a function of angle from the source center
in arcmin for redshift values of $z=[7.5,10,12.5,15]$ for a source
with $\mu = 10^3$ corresponding to a central black hole mass of $10^8
M_\odot$.  The right panels are similar to the left one except that
here the redshift is fixed to z=7.5 and the various curves reflect
values of $\mu=1,10,100,1000$.  The ionisation front convolution
corresponding to a Stromgren sphere, \eg\ sharp ionisation front
appropriate for (stellar) thermal sources, at z=7.5 is shown as well.
}
\label{fig:fig2}
\end{figure*}

Here we have convolved the predicted brightness temperature
fluctuations with the LOFAR beam, taking a Gaussian beam of FWHM 3
arcmin, and obtained the results shown in figure~\ref{fig:fig2}. This
figure shows the expected LOFAR observed differential brightness
temperature for various values of redshifts with fixed central black
hole mass of $10^8 M_\odot$ (left panel) and for different values of
black hole masses with fixed redshift (right panel). It is clear that
the ionisation front for the $\mu=100-1000$ case is larger than the
beam FWHM and could be easily observed with LOFAR. However, this is
not the case for a miniquasar with central black hole mass $\lsim
10^6~M_\odot$. The LOFAR baseline will allow higher resolution
observations but for those baselines the collecting area of the
telescope with respect to the core increases very little, \eg\ a
baseline of $5$~km has about 1.2 the core collecting area, resulting
in a sharp decrease in the telescope sensitivity at these
resolutions. To fully resolve such ionisation fronts, one would have to
wait, \eg\ for the Square Kilometer Array which for the same baseline
has an order of magnitude higher sensitivity than LOFAR.  The extended
wings of the ionised region shown in figures~\ref{fig:fig1} \&
~\ref{fig:fig2} could also be observed with LOFAR, possibly even for
the low power miniquasars. These correspond to a small but not
negligible ionised fraction ($x_{HI}\gsim 0.8$) with a $\delta T_b$ of
the order of a few. The extended region of these low ionisation wings
compensates for their low signal and might possibly render them
observable with LOFAR.

\section{summary}

In this paper we have shown that for very simple assumptions,
LOFAR could in principle indicate whether a certain ionisation front
is caused by miniquasars or stars by measuring its width. Our
assumptions include a single spherical symmetric ionisation source
with no overlapping ionised spheres, a single ionising  population
and a homogeneous background with no clustering. In the future, we will
include the source clustering effects and the effects of the source
luminosity function, and compute the LOFAR correlation function.

In reality, however, the reionisation process might have been much
more complex and have been induced by multiple types of ionisation
radiation sources that were dominant at various stages of the universe
history and present in an inhomogeneous medium.  For example, massive
Population III stars might die early on and create miniquasars that can
maintain the IGM to be partially ionised through their hard photons
(Ricotti \& Ostriker 2004a; 2004b; Madau \etal\ 2004) while the full
ionisation could be attributed to the soft miniquasar photons at later
stages.  Notwithstanding the complex ionisation history of the IGM,
the thickness of the ionisation fronts expected to be observed via the
21 cm experiments \eg\, LOFAR and PAST (Peterson, Pen and Wu, 2004)
should still be a very strong discriminator between stellar and
accreting black hole ionising radiation sources.

\section*{acknowledgments}
J. S. acknowledges the hospitality of the Kapteyn Astronomical
Institute. The authors thank A.G. de Bruyn and P. Madau for
discussions.

{}
 
\end{document}